\documentclass{appolb}
\usepackage{graphicx}
\usepackage{multirow}
\usepackage{ctable}
\usepackage{booktabs}
\usepackage{array}
\usepackage{tabularx}
\usepackage{xcolor}
\usepackage{pstricks}

\begin{document}
\title{$D \bar D$ asymmetry at low and high energies and 
possible consequences for prompt atmospheric neutrinos%
\thanks{Presented at XXIV Cracow EPIPHANY Conference on Advances in Heavy Flavour Physics, 9-12 January 2018}%
}
\author{Antoni Szczurek, Rafa{\l} Maciu{\l}a
\address{Institute of Nuclear
Physics, Polish Academy of Sciences, Radzikowskiego 152, PL-31-342 Krak{\'o}w, Poland}
}
\maketitle
\begin{abstract}
We discuss the role of unfavoured light quark/antiquark to $D$ meson 
fragmentation. The unknown parameters of fragmentation process
are adjusted to describe the asymmetry for $D^+$ and $D^-$ production 
measured by the LHCb. 
Predictions for similar asymmetry for neutral $D$ mesons are presented.
The predicted asymmetry at large rapidity (or $x_F$) are very large
which is related to the valence-quark contribution. 
As a result, prompt atmospheric neutrino flux at high neutrino energies 
can be much larger than for the conventional $c \to D$ fragmentation. 
We predict large rapidity-dependent $D^+/D^-$ and $D^0/{\bar D}^0$ 
asymmetries for low ($\sqrt{s} =$ 20 - 100 GeV) energies. 
The $q/\bar q \to D$ fragmentation leads to enhanced 
production of $D$ mesons at low energies.
Predictions for fixed target $p+^{4}\!\textrm{He}$ collisions relevant 
for a fixed target LHCb experiment are discussed.
\end{abstract}
  
\section{Introduction}

It is believed that the high-energy neutrinos observed by 
the IceCube collaboration (see e.g. \cite{IceCube_flux}) are 
of extraterrestial origin.
Another important component comes from semileptonic decays 
of $D$ mesons produced in the atmosphere by the collision of cosmic 
rays (mostly protons) with the atmosphere (mostly $^{14}N$).
The flux of cosmic rays (charged particles) is relatively well known
as measured e.g. by the Auger experiment.
It is well known that the dominant mechanism of charm production
at high energies is $g g \to c \bar c$ partonic subprocesses.

Recently we have performed a critical analysis of uncertainties
in the high-energy production of charm (D mesons) \cite{GMPS2017}.
The following conclusions were obtained in \cite{GMPS2017}.
The high-energy neutrinos are produced mostly in very high-energy
proton-proton collisions (larger than at the LHC).
The region of $x_F >$ 0.3 is crucial for high-energy neutrinos, 
however not accessible at the LHC.
Both very small and very large longitudinal momentum fractions of gluons
are important. These regions are not well known.

Recently the LHCb collaboration observed $D^+$ and $D^-$ asymmetry 
at forward directions \cite{LHCb_asymmetry}.
In the literature one routinely assumes that D mesons are produced from 
$c$ or $\bar c$ fragmentation. This gives no asymmetry, 
at least in leading order approach.

Recently we have considered also subleading unfavoured fragmentation 
\cite{MS2018}.
It is known that the unfavoured fragmentation leads to 
asymmetry in $K^+$ and $K^-$ production (SPS, RHIC/BRAHMS).
Also $\pi^+ \pi^-$ asymmetry was observed but there both quark and
antiquark fragmentation functions are (assumed) the same.

\section{A sketch of our approach}

\subsection{Unfavoured fragmentation}

The dominant at large $x_F$ high-energy processes:
$u g \to u g$, $d g \to d g$, $\bar u g \to \bar u g$ and 
$\bar d g \to \bar d g$ and subsequent light quark/antiquark to 
D meson fragmentation and/or decays are calculated in 
the leading-order (LO) collinear factorization approach with a special
treatment of minijets at low transverse momenta,
as adopted in \textsc{Pythia}, 
by multiplying standard cross section by a somewhat arbitrary
suppression factor:
\begin{equation}
F_{sup}(p_T) = \frac{p_T^4}{((p_{T}^{0})^{2} + p_T^2)^2} \theta(p_T -
p_{T,cut}) \; .
\label{suppression_factor}
\end{equation}

\begin{figure}[!h]
\centering
\begin{minipage}{0.47\textwidth}
 \centerline{\includegraphics[width=1.0\textwidth]{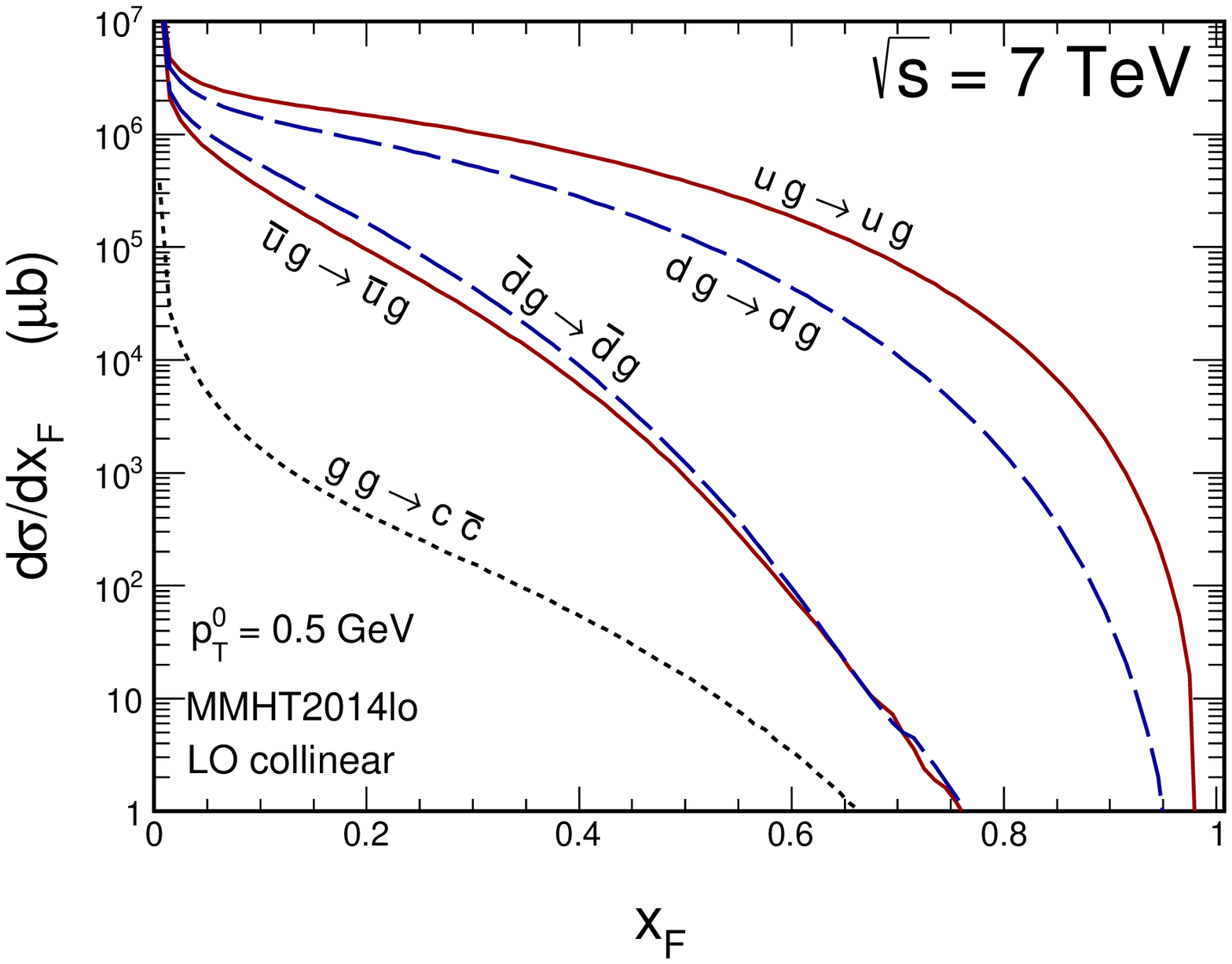}}
\end{minipage}
\hspace{0.5cm}
\begin{minipage}{0.47\textwidth}
 \centerline{\includegraphics[width=1.0\textwidth]{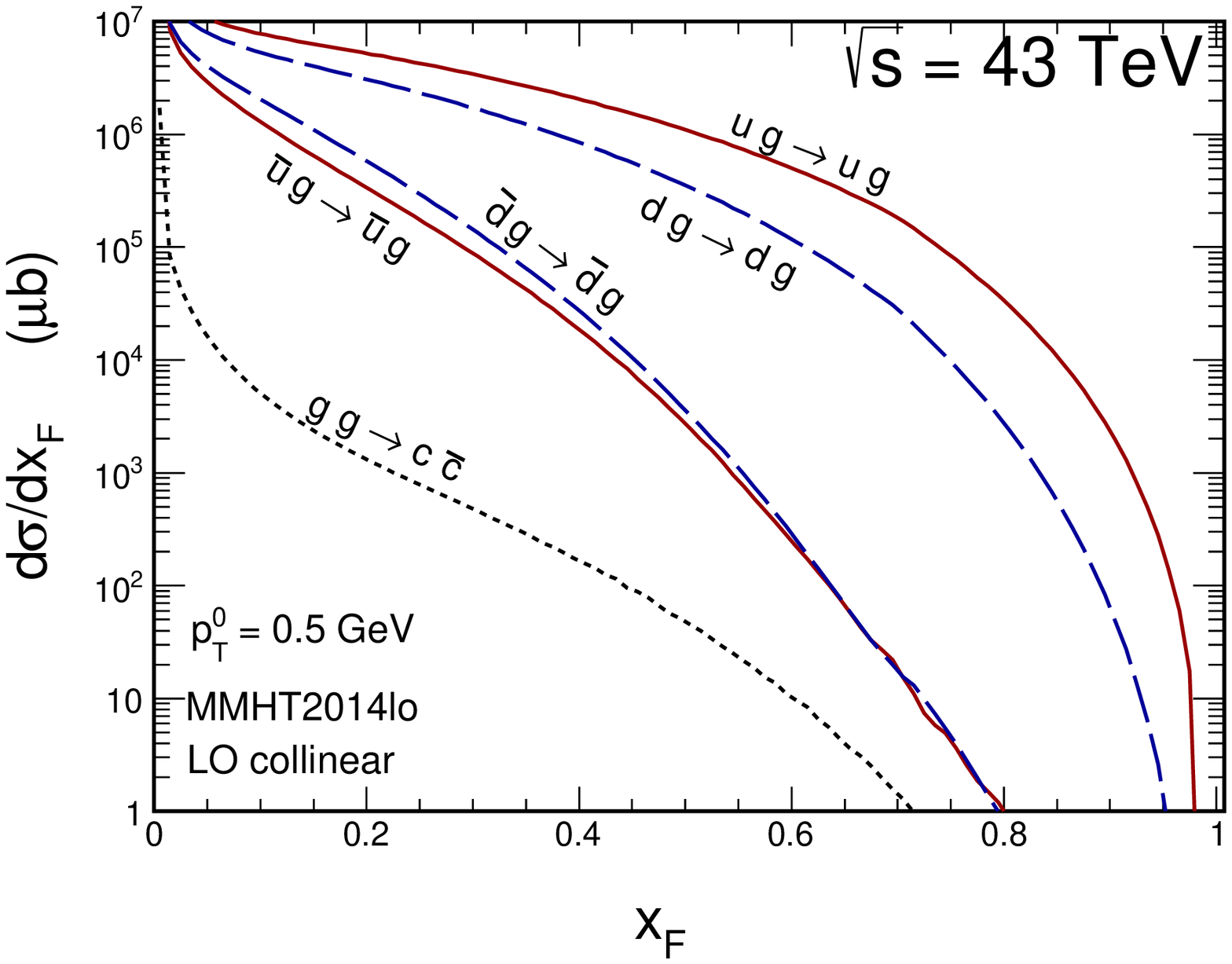}}
\end{minipage}
   \caption{
\small Quark and antiquark distributions in Feynman $x_F$ for 
$\sqrt{s} =$ 7 TeV (left panel) and $\sqrt{s} =$ 43 TeV (right panel)
corresponding to $E_{\mathrm{lab}}(p)$ = 10$^{9}$ GeV 
(relevant for high-energy prompt atmospheric neutrinos).
This calculation was performed within collinear-factorization approach
with $p_{T}^{0} =$ 0.5 GeV.
 }
 \label{fig:dsig_dxf_partons}
\end{figure}

In Fig.\ref{fig:dsig_dxf_partons} we show distributions of
quarks/aniquarks produced in such mechanisms.
In this leading-order calculation we have used the regulator given by 
Eq.(\ref{suppression_factor}).
We can observe much larger cross section than for $c/{\bar c}$
production in the region of large $x_F$.

To get distributions of mesons we have to include 
$u,\bar u, d, \bar d \to D^i$ parton fragmentation. 
The corresponding fragmentation functions fulfill the following
flavour symmetry conditions:
\begin{equation}
D_{d \to D^-}(z) = D_{\bar d \to D^+}(z) = D^{(0)}(z) \; .
\label{ff_symmetries}
\end{equation}
Similar symmetry relations hold for fragmentation 
of $u$ and $\bar u$ to $D^0$ and $\bar D^0$ mesons.
However $D_{q \to D^0}(z) \ne D_{q \to D^+}(z)$ which is caused
by the contributions from decays of vector $D^*$ mesons.
Furthermore we assume:
\begin{equation}
D_{\bar u \to D^{\pm}}(z) = D_{u \to D^{\pm}}(z) = 0 \; .
\label{neglected_ff}
\end{equation}
for doubly suppressed fragmentations.

We limit in the following to a phenomenological approach
and ignore possible DGLAP evolution effects important at somewhat 
larger transverse momenta.
We parametrize the unfavoured fragmentation functions in 
the low-$p_t$ phase space region as:
\begin{equation}
D_{q \to D}(z) = A_{\alpha} (1-z)^{\alpha} \; .
\label{ff_simple_parametrization}
\end{equation}
Instead of fixing the uknown $A_{\alpha}$ we will operate rather with
the fragmentation probability:
\begin{equation}
P_{q \to D} = \int dz \; A_{\alpha} \left( 1 - z \right)^{\alpha} \; .
\label{Dff_simple_parametrization}
\end{equation}
and calculate corresponding $A_{\alpha}$ for a fixed $P_{q \to D}$ and
$\alpha$.
In our approach we have only two free parameters.

Another simple option we considered in \cite{MS2018} is:
\begin{equation}
D_{q_f \to D}(z) = P_{q_f \to D} \cdot D_{\mathrm{Peterson}}(1-z) \; .
\label{Peterson}
\end{equation}
For heavy quark fragmentation ($c \to D$) 
the Peterson fragmentation function is peaked at large $z$. 
The light quark/antiquark fragmentation is expected to be dominant at small
$z$. This is the case of Peterson fragmentation function reflected
with respect to $z$ = 1/2. We used such a purely phenomenological
function as another example to test uncertainties related to the shape of
the a priori unknown function.

In addition to the direct fragmentation (given by $D^{(0)}(z)$) 
there are also contributions with intermediate vector $D^*$ mesons.
Then the chain of production of charged $D$ mesons is as follows:
\begin{eqnarray}
&&\bar u \to D^{*,0} \to D^+ \; \mathrm{(forbidden)}, \nonumber \\
&&     u \to {\bar D}^{*,0} \to D^- \; \mathrm{(forbidden)}, \nonumber \\
&&\bar d \to D^{*,+} \to D^+ \; \mathrm{(allowed)}, \nonumber \\
&&     d \to D^{*,-} \to D^- \; \mathrm{(allowed)}.
\label{intermediate_vectors}  
\end{eqnarray}
Including both direct and feed-down contributions the combined 
fragmentation function of light quarks/antiquarks to charged $D$ mesons 
can be written as:
\begin{equation}
D_{d/\bar d \to D^{\mp}}^{\mathrm{eff}}(z) =
D_{d/\bar d \to D^{\mp}}^0(z) +
P_{\mp \to \mp} \cdot D_{d/\bar d \to D^{*,\mp}}^1(z) \; .
\label{charged_D_mesons}
\end{equation}
Similar formula can be written for neutral $D$ mesons \cite{MS2018}.
We assume flavour symmetry of fragmentation functions also 
for vector $D$ meson production.
In our calculations in \cite{MS2018} we assumed in addition:
\begin{equation}
D^{(0)}(z) \approx D^{(1)}(z) 
\end{equation}
which can be easily modified if needed.

\subsection{Production asymmetry}

The flavour asymmetry in production is defined as:
\begin{equation}
A_{D^+/D^-}(\xi) 
= \frac{ \frac{d \sigma_{D^-}}{d \xi}(\xi) - \frac{d \sigma_{D^+}}{d \xi}(\xi) }
       { \frac{d \sigma_{D^-}}{d \xi}(\xi) + \frac{d \sigma_{D^-}}{d \xi}(\xi) }
\; ,
\label{asymmetry}
\end{equation}
where $\xi = x_F, y, p_T, (y,p_T)$.
In \cite{MS2018} we have considered several examples.

\section{Results}

\subsection{LHCb asymmetry}

\begin{figure}[!h]
\begin{minipage}{0.47\textwidth}
  \centerline{\includegraphics[width=1.0\textwidth]{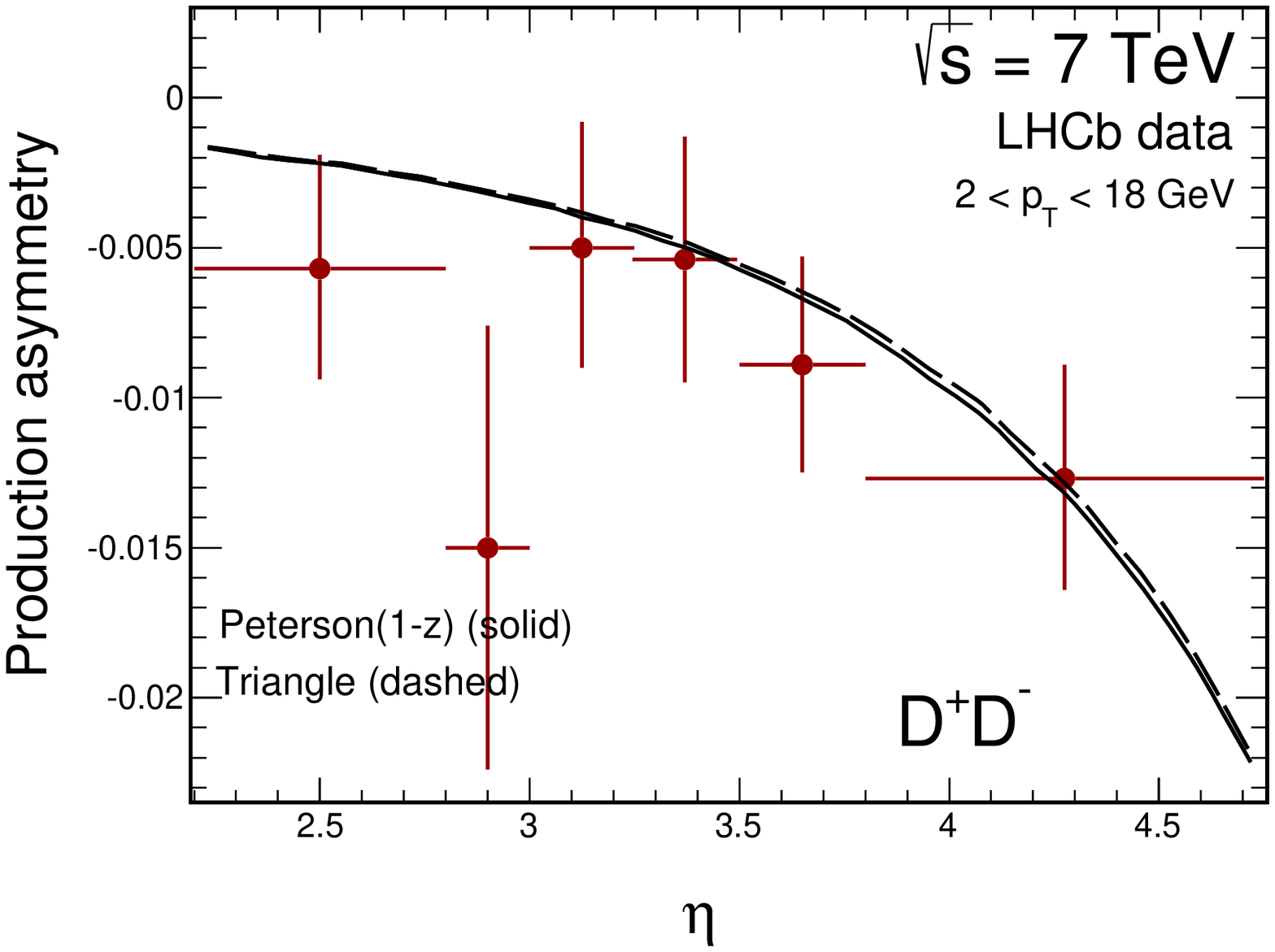}}
\end{minipage}
\hspace{0.5cm}
\begin{minipage}{0.47\textwidth}
  \centerline{\includegraphics[width=1.0\textwidth]{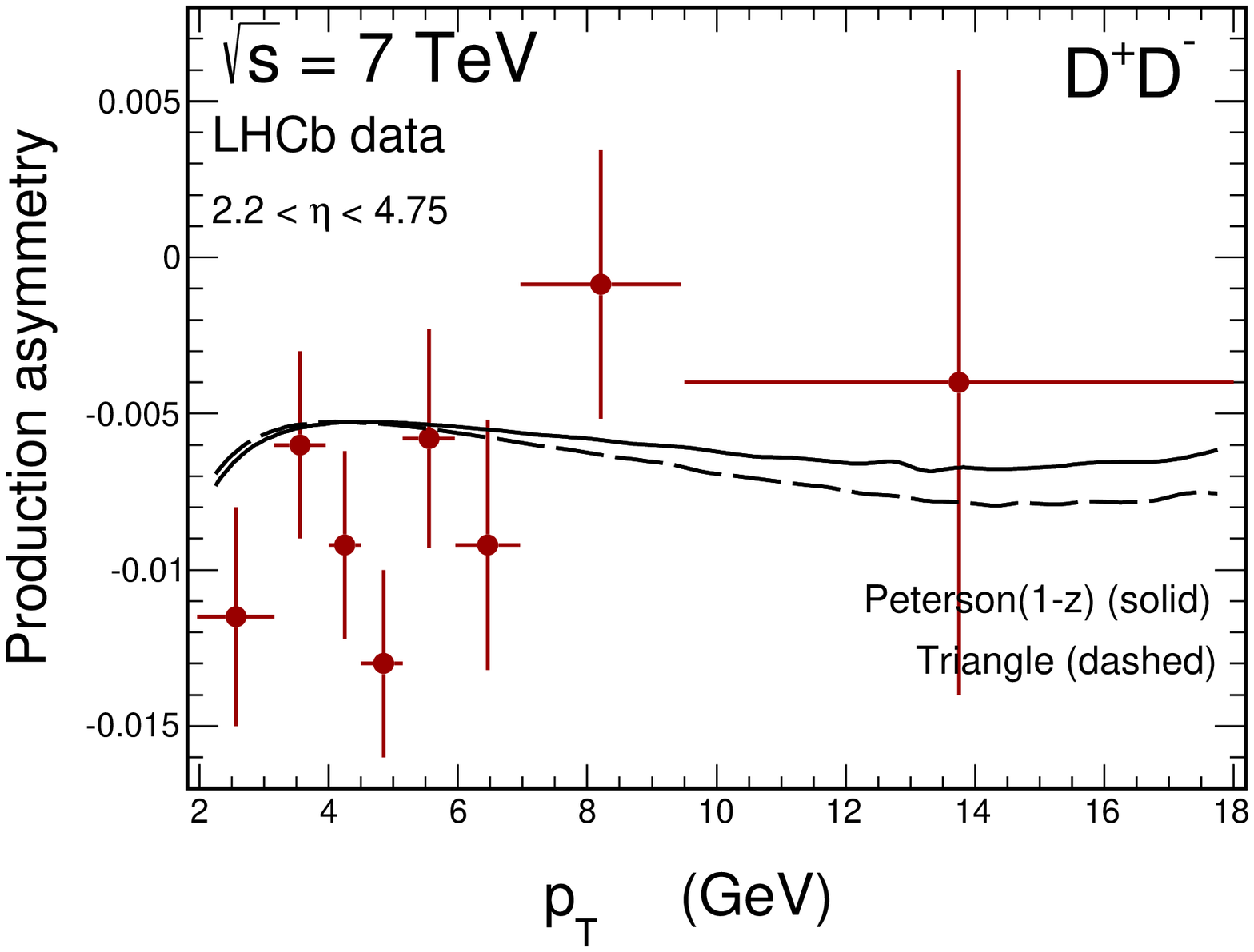}}
\end{minipage}
\begin{minipage}{0.47\textwidth}
  \centerline{\includegraphics[width=1.0\textwidth]{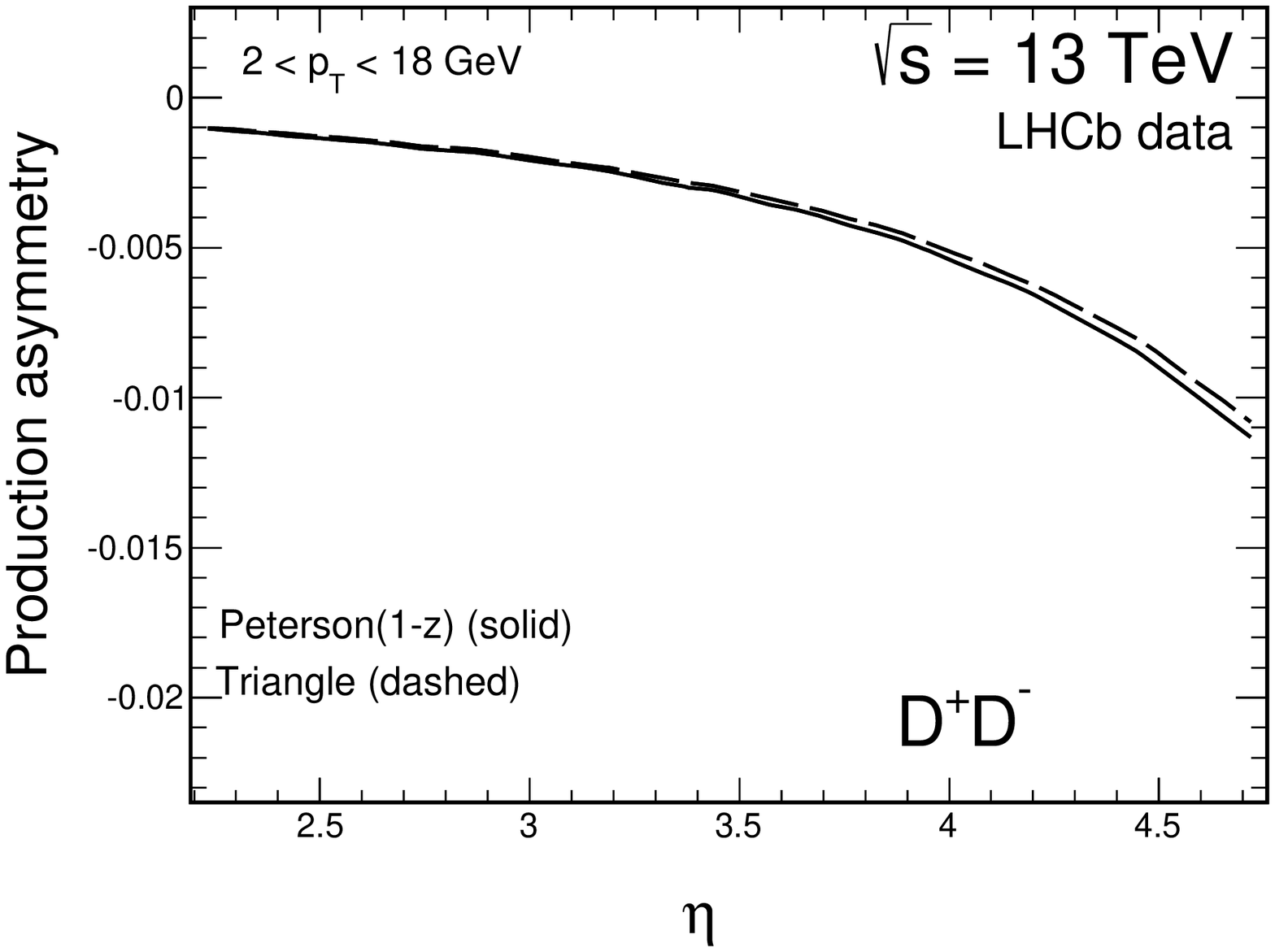}}
\end{minipage}
\hspace{0.5cm}
\begin{minipage}{0.47\textwidth}
  \centerline{\includegraphics[width=1.0\textwidth]{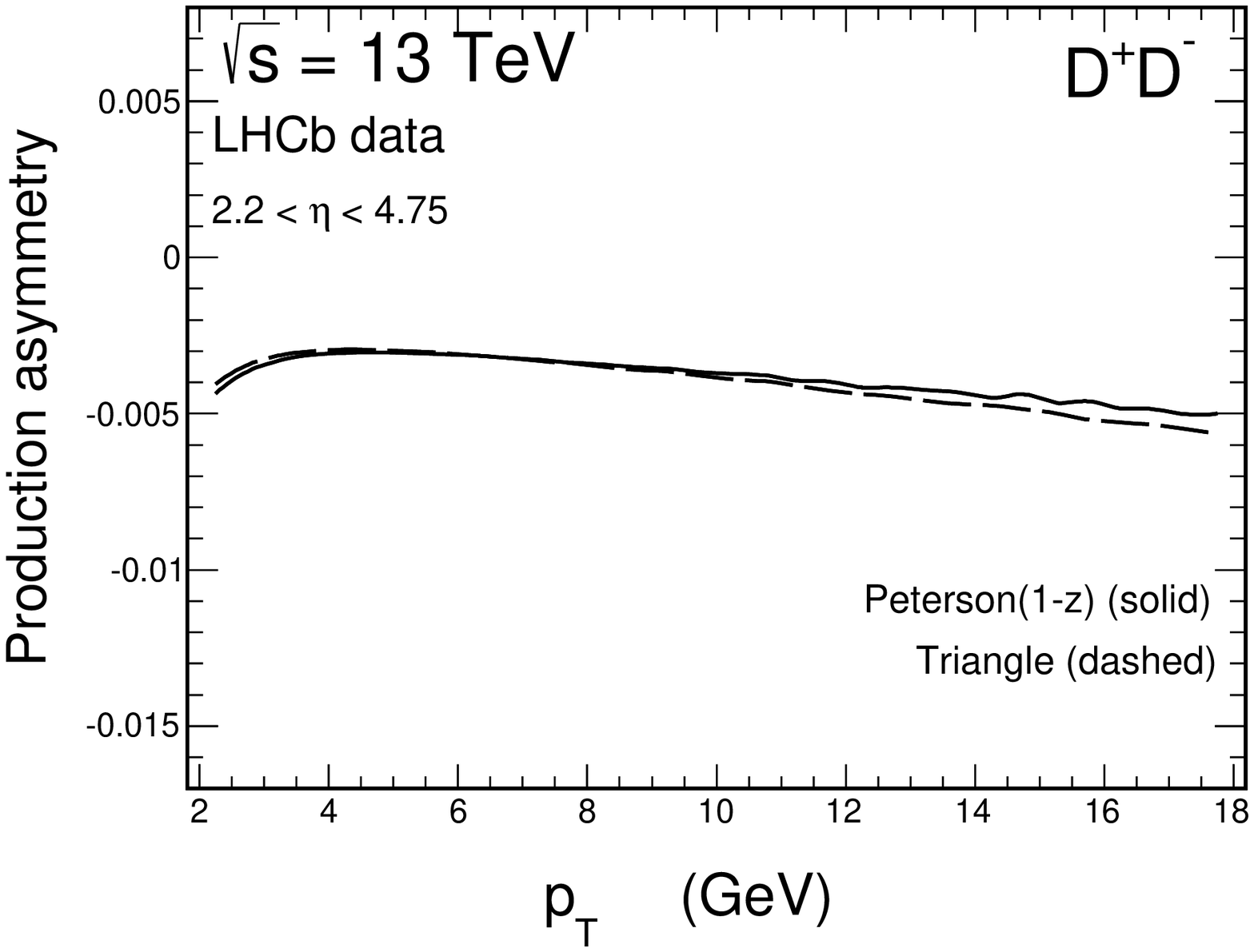}}
\end{minipage}
  \caption{
\small $A_{D^+/D^-}$ production asymmetry measured by the LHCb
collaboration at $\sqrt{s}= 7$ TeV as a function of $D$ meson
pseudorapidity (left-top panel) and $D$ meson
transverse momentum (right-top panel). 
The corresponding predictions for $\sqrt{s}= 13$ TeV are shown 
in the bottom panels.
}
\label{fig:LHCb_asymmetry_charged}
\end{figure}

In the top panels of Fig.~\ref{fig:LHCb_asymmetry_charged} we 
show results for the asymmetry for $P_{q \to D}$ adjusted to the LHCb
data. In this calculation we have fixed $\alpha$ = 1 in formula 
(\ref{ff_simple_parametrization}).
In the left panel we show $A_{D^+/D^-}(\eta)$ for $p_{T,D} \in$ (2,18) GeV
and in the right panel we show $A_{D^+/D^-}(p_T)$ for 2.2 $< \eta <$ 4.75 .
We find that $P_{q \to D} =$ 0.005 $\pm$ 0.001 for triangle
fragmentation function 
and $P_{q \to D} =$ 0.007 $\pm$ 0.001 for Peterson(1-z) is consistent with 
main trends of the LHCb data. This are rather small numbers compared
to $c/{\bar c} \to D/{\bar D}$ fragmentation which happens with probability
of the order of 50 \%. The results only weakly depend on transverse momentum
cut $p_{T}^0$, since the LHCb kinematics excludes the uncertain region 
of very small meson transverse momenta. 
In the bottom panels we show our predictions for $\sqrt{s}=13$ TeV.

In \cite{MS2018} we showed also our predictions for $D^0-{\bar D}^0$
asymmetry.

\subsection{Low energies}

The discussed by us mechanisms of subleading fragmentation of $D$ mesons
lead to enhanced production of $D$ mesons at lower energies.
In Table~\ref{tab:low_energies} we show different
contributions to the production of $D^{+}/D^{-}$ mesons. The dominant at
high-energy $gg \to c \bar c$ mechanism gives only $13\%$ and $18\%$ 
for $\sqrt{s}=27$ and $39$ GeV, respectively and strongly underestimates the
NA27 \cite{AguilarBenitez:1988sb} and E743 \cite{Ammar:1988ta}
experimental data. Inclusion of the "subleading" contributions brings
theoretical calculations much closer to the experimental data.
We predict sizeable $D^{+}/D^{-}$ asymmetries at these low energies. 

\begin{table}[tb]%
\caption{Different contributions to the cross sections (in microbarns) 
for $D^{+}+D^{-}$ production at low energies. The results presented here
were obtained with $p_{T}^{0} = 1.5$ GeV.}

\label{tab:low_energies}
\centering %
\resizebox{\textwidth}{!}{%
\begin{tabularx}{14.5cm}{c c c}
\\[-2.ex] 
\toprule[0.1em] %
\\[-2.ex] 

\multirow{1}{4.5cm}{process:} & \multirow{1}{3.cm}{$\sqrt{s}=27$ GeV} & \multirow{1}{3.cm}{$\sqrt{s}=39$ GeV}\\ [+0.1ex]
\bottomrule[0.1em]
\multirow{1}{4.5cm}{$g^* g^* \to c\bar{c} \;\;\; (c/\bar{c} \to D^{\pm})$} & \multirow{1}{3.cm}{$1.52$} & \multirow{1}{3.cm}{$4.58$}\\ [+0.1ex]
\multirow{1}{4.5cm}{$q^* \bar q^* \to c\bar{c} \;\;\; (c/\bar{c} \to D^{\pm})$} & \multirow{1}{3.cm}{$0.08$} & \multirow{1}{3.cm}{$0.19$}\\ [+0.1ex]
\hline
\multirow{1}{4.5cm}{$g d \to g d \;\;\;\; (d \to D^{-})$} & \multirow{1}{3.cm}{$9.53$} & \multirow{1}{3.cm}{$13.89$}\\ [-0.2ex]
\multirow{1}{4.5cm}{$g \bar{d} \to g \bar{d} \;\;\;\; (\bar{d} \to D^{+})$} & \multirow{1}{3.cm}{$3.03$} & \multirow{1}{3.cm}{$4.78$}\\ [+0.1ex]
\hline
\multirow{1}{4.5cm}{$d d \to d d \;\;\;\; (d \to D^{-}) \times 2$} & \multirow{1}{3.cm}{$3.07$} & \multirow{1}{3.cm}{$4.29$}\\ [-0.2ex]
\multirow{1}{4.5cm}{$\bar{d} \bar{d} \to \bar{d} \bar{d} \;\;\;\; (\bar{d} \to D^{+}) \times 2$} & \multirow{1}{3.cm}{$0.29$} & \multirow{1}{3.cm}{$0.49$}\\ [-0.2ex]
\hline
\multirow{1}{4.5cm}{$\bar{d} d \to \bar{d} d  \;\;\;\; (d \to D^{-})$} & \multirow{1}{3.cm}{$0.58$} & \multirow{1}{3.cm}{$0.88$}\\ [-0.2ex]
\multirow{1}{4.5cm}{$d \bar{d} \to d \bar{d} \;\;\;\; (\bar{d} \to D^{+})$} & \multirow{1}{3.cm}{$0.58$} & \multirow{1}{3.cm}{$0.88$}\\ [-0.2ex]
\hline
\multirow{1}{4.5cm}{$u d \to u d \;\;\;\; (d \to D^{-})$} & \multirow{1}{3.cm}{$2.76$} & \multirow{1}{3.cm}{$3.72$}\\ [-0.2ex]
\multirow{1}{4.5cm}{$\bar{u} \bar{d} \to \bar{u} \bar{d} \;\;\;\; (\bar{d} \to D^{+})$} & \multirow{1}{3.cm}{$0.12$} & \multirow{1}{3.cm}{$0.19$}\\ [-0.2ex]
\hline
\multirow{1}{4.5cm}{$\bar{u} d \to \bar{u} d \;\;\;\; (d \to D^{-})$} & \multirow{1}{3.cm}{$0.40$} & \multirow{1}{3.cm}{$0.63$}\\ [-0.2ex]
\multirow{1}{4.5cm}{$u \bar{d} \to u \bar{d} \;\;\;\; (\bar{d} \to D^{+})$} & \multirow{1}{3.cm}{$0.97$} & \multirow{1}{3.cm}{$1.42$}\\ [-0.2ex]
\bottomrule[0.1em]
\multirow{1}{4.5cm}{theory predictions} & \multirow{1}{3.cm}{$22.93$} & \multirow{1}{3.cm}{$35.94$}\\ [-0.2ex]
\bottomrule[0.1em]
\multirow{1}{4.5cm}{experiment} & \multirow{1}{5.5cm}{NA27: $11.9 \pm 1.5$} & \multirow{1}{5.cm}{E743: $26 \pm 4 \pm 25\%$}\\ [-0.2ex]
\bottomrule[0.1em]
\end{tabularx}
}
\end{table}

\begin{figure}[!h]
\centering
\begin{minipage}{0.47\textwidth}
  \centerline{\includegraphics[width=1.0\textwidth]{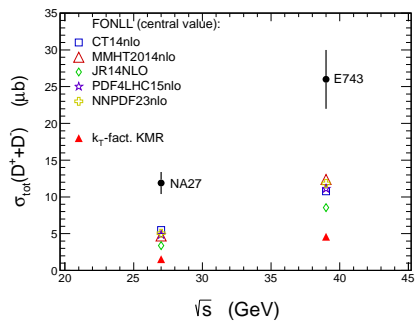}}
\end{minipage}
  \caption{
\small Total cross section for $D^+ + D^-$ production.
The experimental data are from Refs.~\cite{AguilarBenitez:1988sb} 
and \cite{Ammar:1988ta}.
The details of different calculations are explained in the figure.
}
\label{fig:FONLL}
\end{figure}

In Fig.~\ref{fig:FONLL} we show the lowest energy data for 
charged $D$ mesons in proton-proton collisions 
\cite{AguilarBenitez:1988sb,Ammar:1988ta}. 
We show results of conventional calculation in the $k_T$-factorization 
as well as results obtained with the code FONLL. 
Both the $k_T$-factorization as well as FONLL results are 
below experimental data extrapolated to the full phase
space. Including also theoretical uncertainties, this leaves room for
our subleading fragmentation contribution.
In our paper it was obtained by extrapolating our results,
assuming some parametrizations of the subleading
fragmentation function, to low energies based on the asymmetry 
measured by the LHCb collaboration. Of course our estimate
of the LHCb asymmetry as well as extrapolation
to other corners of the phase space cannot be too precise.
Clearly better data for intermediate and low energies are needed
to constrain the subleading fragmentation. 

\begin{figure}[!h]
\centering
\begin{minipage}{0.47\textwidth}
  \centerline{\includegraphics[width=1.0\textwidth]{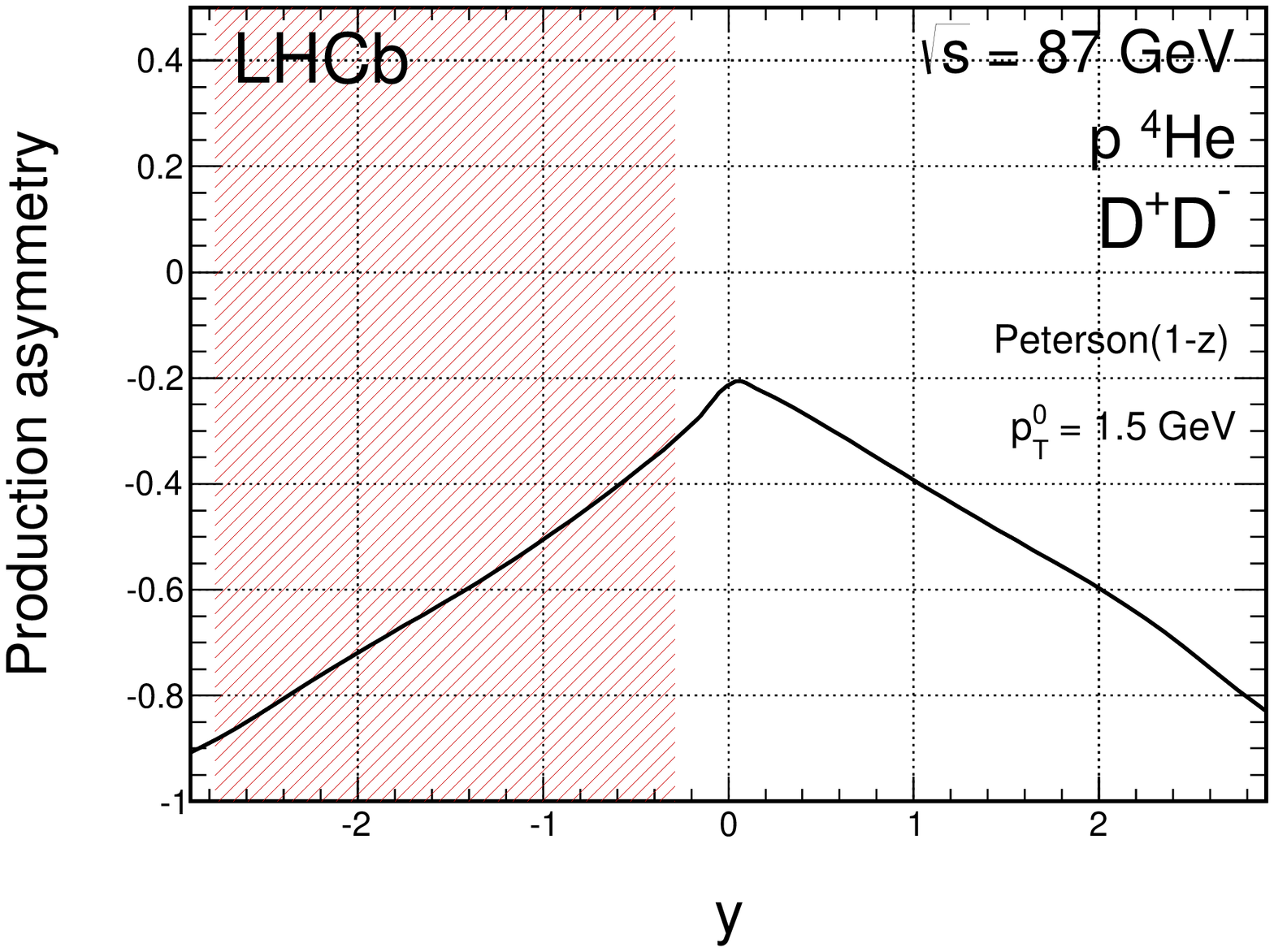}}
\end{minipage}
  \caption{
\small $A_{D^{+}D^{-}}(y)$ production asymmetry for the fixed target
$p+^4\!He$ reaction for $\sqrt{s}=87$ GeV.
}
\label{fig:assym_LHCb_fixed_target}
\end{figure}

The LHCb collaboration has good experience in measuring the asymmetry
in $D^+$ and $D^-$ production. 
Such an analysis can be done e.g. for fixed target experiment 
$p + ^{4}\!\mathrm{He}$ with gaseous target.
The nuclear effects for $^{4}$He are rather small.
Neglecting the nuclear effects the differential cross section  
for production of $q/\bar q$ (particle 1) and 
associated parton (particle 2) can be written
in the collinear factorization approach as:
\begin{equation}
  \frac{d \sigma_{p\; ^4\! \mathrm{He}}}{d y_1 d y_2 d p_{T}} =
2 \frac{d \sigma_{pp}}{d y_1 d y_2 d p_{T}} +
2 \frac{d \sigma_{pn}}{d y_1 d y_2 d p_{T}} \; .
\label{nuclear_cross_section} 
\end{equation}
In Fig.~\ref{fig:assym_LHCb_fixed_target} we present the relevant
predictions for the LHCb experiment.
Rather large asymmetries are predicted which could be addressed in the
expected analysis of the fixed target experiment.

\begin{figure}[!h]
\centering
\begin{minipage}{0.47\textwidth}
  \centerline{\includegraphics[width=1.0\textwidth]{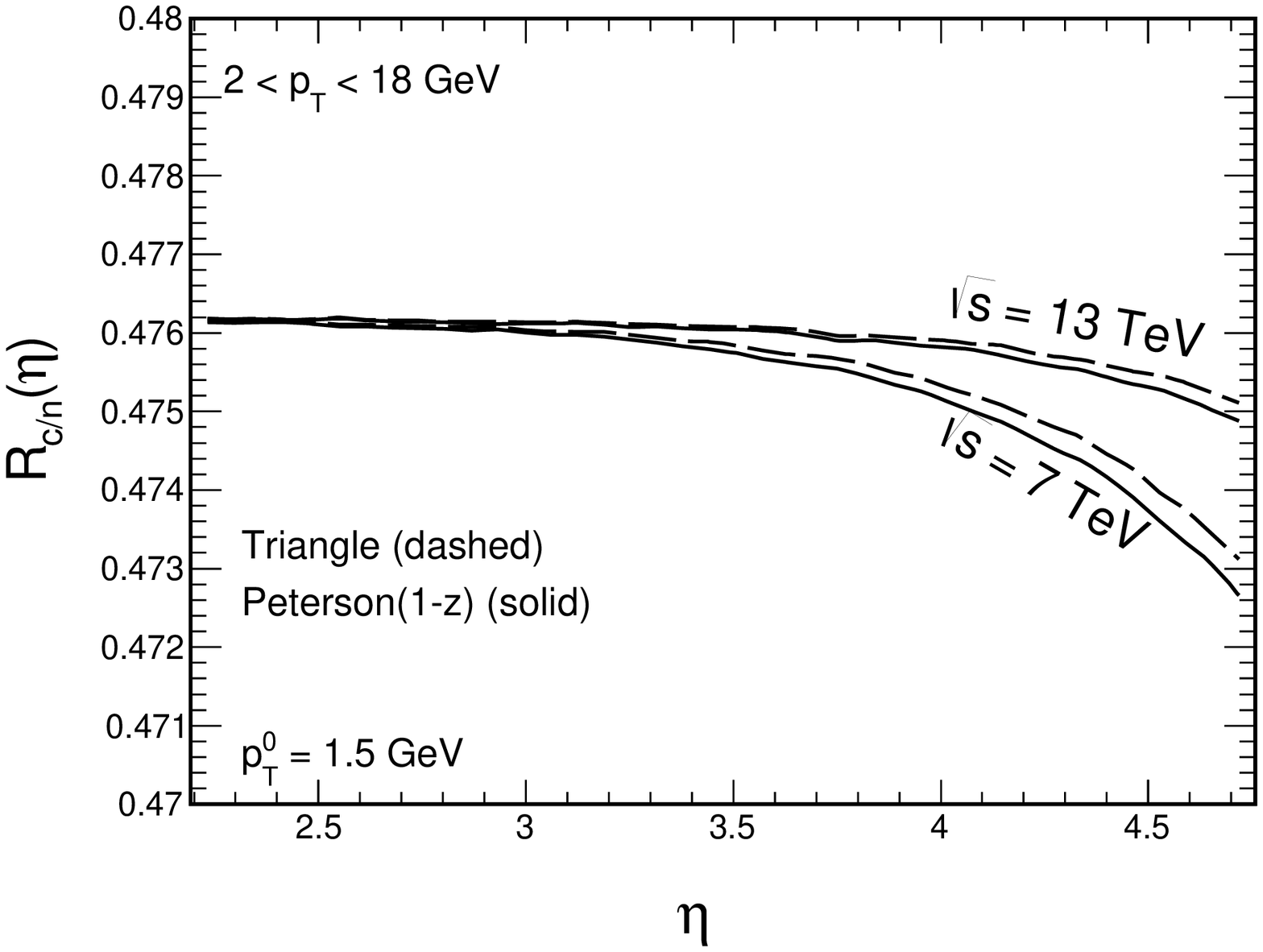}}
\end{minipage}
\begin{minipage}{0.47\textwidth}
  \centerline{\includegraphics[width=1.0\textwidth]{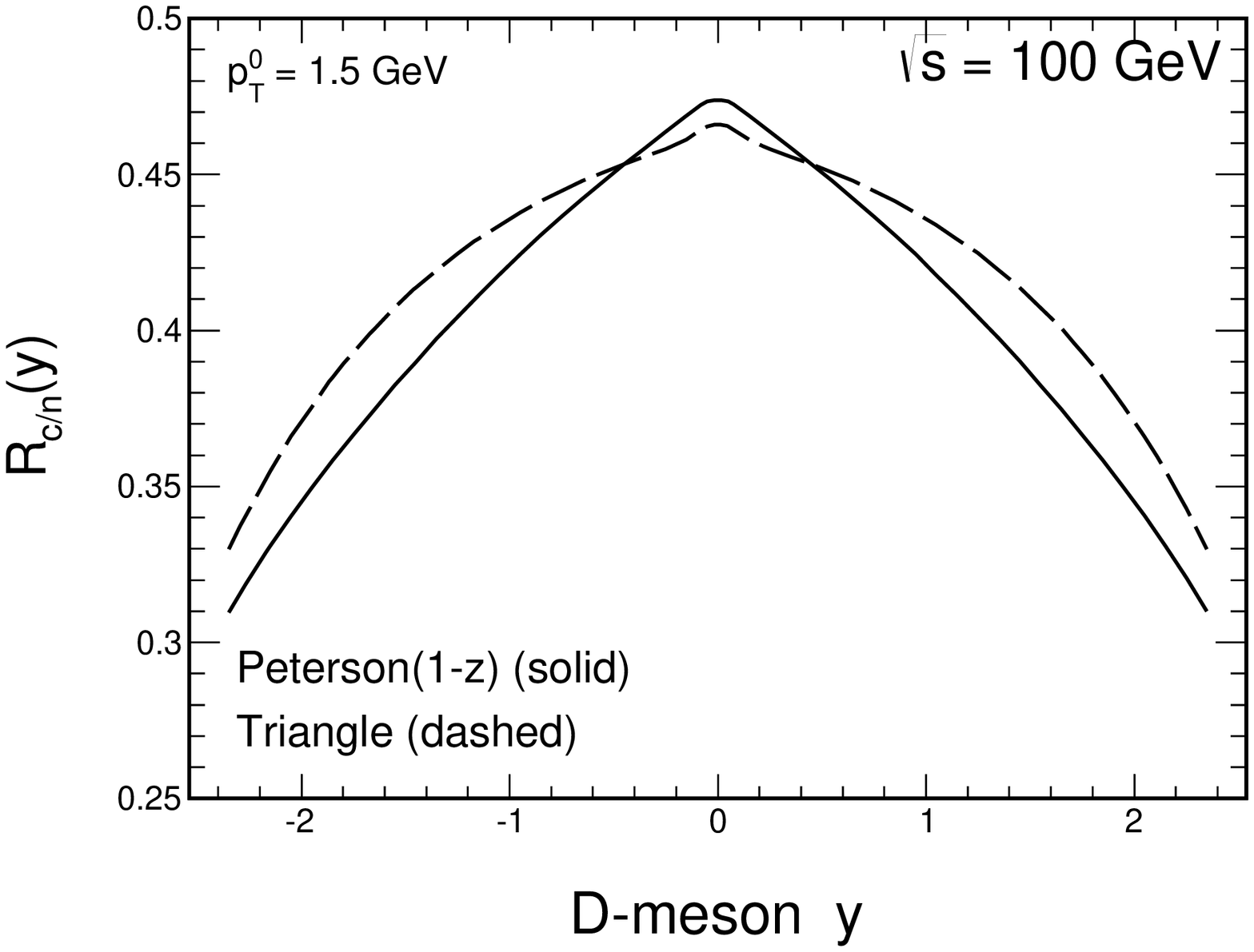}}
\end{minipage}
  \caption{
\small The $R_{c/n}$ ratio as a function of meson pseudorapidity for 
$\sqrt{s}= 7$ and $13$ TeV for the LHCb kinematics (left panel) and
as a function of meson rapidity for $\sqrt{s}$ = 100 GeV 
in the full phase-space (right panel). 
Only quark-gluon subleading components are included here.
}
\label{fig:R_cton}
\end{figure}

In the traditional pQCD approach (production of $c/\bar c$ and
only $c/ \bar c \to D / \bar D$ fragmentation) the ratio defined as
\begin{equation}
R_{c/n} \equiv \frac{D^+ + D^-}{D^0 + {\bar D}^0} 
\label{R_cton}
\end{equation}
is a constant, independent of collision energy and rapidity (or $x_F$).
Inclusion of the unfavoured  contribution changes the situation.
In Fig.~\ref{fig:R_cton} we show the ratio as a function of 
meson pseudorapidity $\eta$ for LHC energies (left panel) and 
meson rapidity $y$ for $\sqrt{s}=100$ GeV (right panel), 
taking into account the subleading contribution. At the LHC energies 
very small, difficult to measure, effect is found for the LHCb
transverse momentum and pseudorapidity range. 
At $\sqrt{s}=100$ GeV we predict a strong rapidity dependence of 
the $R_{c/n}$ ratio. We think that fixed target experiments at the LHCb
could address the issue.

\subsection{High energies}

In this subsection we wish to show results relevant for high-energy
prompt atmospheric neutrinos. As discussed recently in Ref.~\cite{GMPS2017} 
a rather large $x_F \sim$ 0.5 region is important in this context.
The $d \sigma / d x_F$ distribution of mesons is the most appropriate 
distribution in this context. 

\begin{figure}[!h]
\begin{minipage}{0.47\textwidth}
  \centerline{\includegraphics[width=1.0\textwidth]{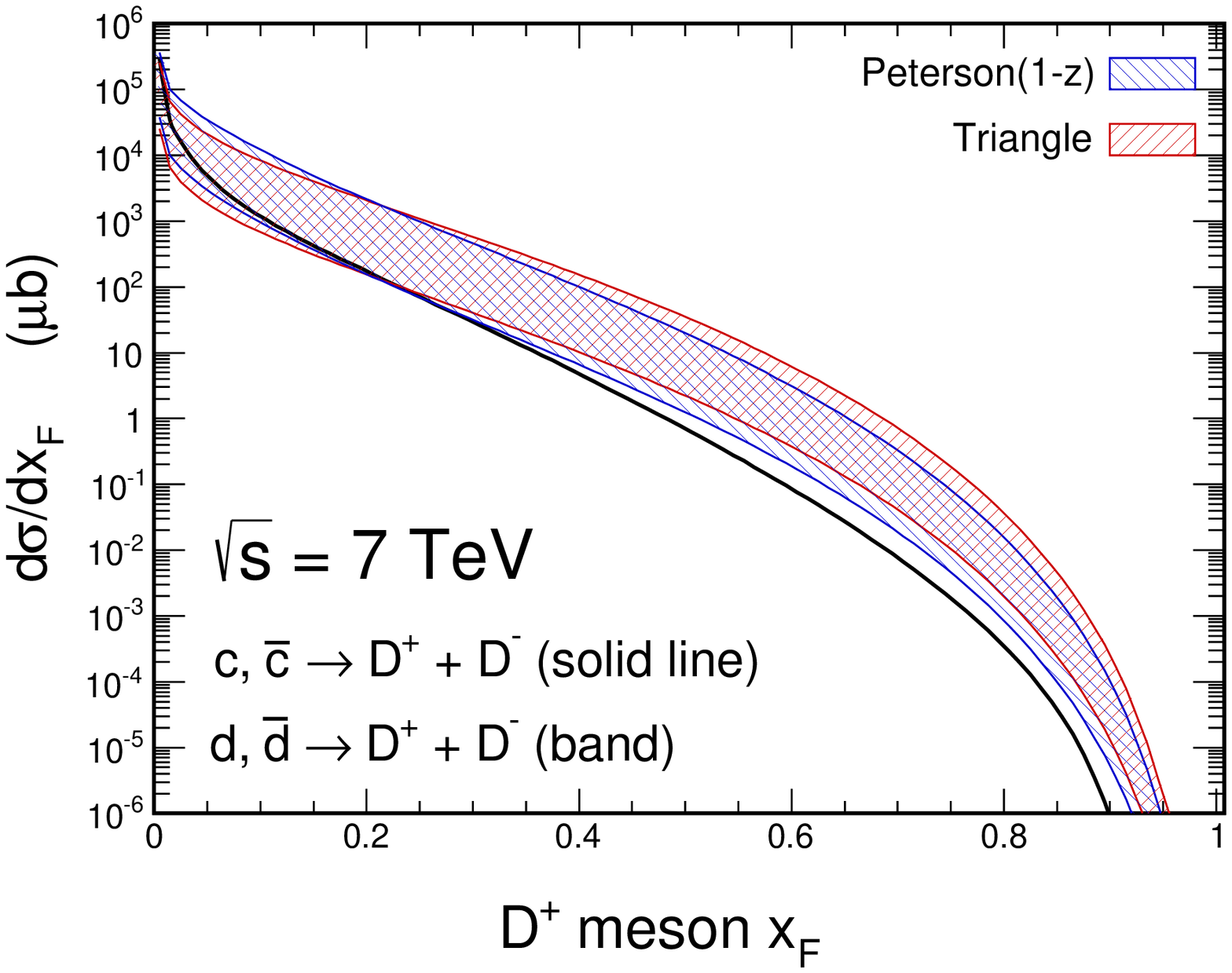}}
\end{minipage}
\hspace{0.5cm}
\begin{minipage}{0.47\textwidth}
  \centerline{\includegraphics[width=1.0\textwidth]{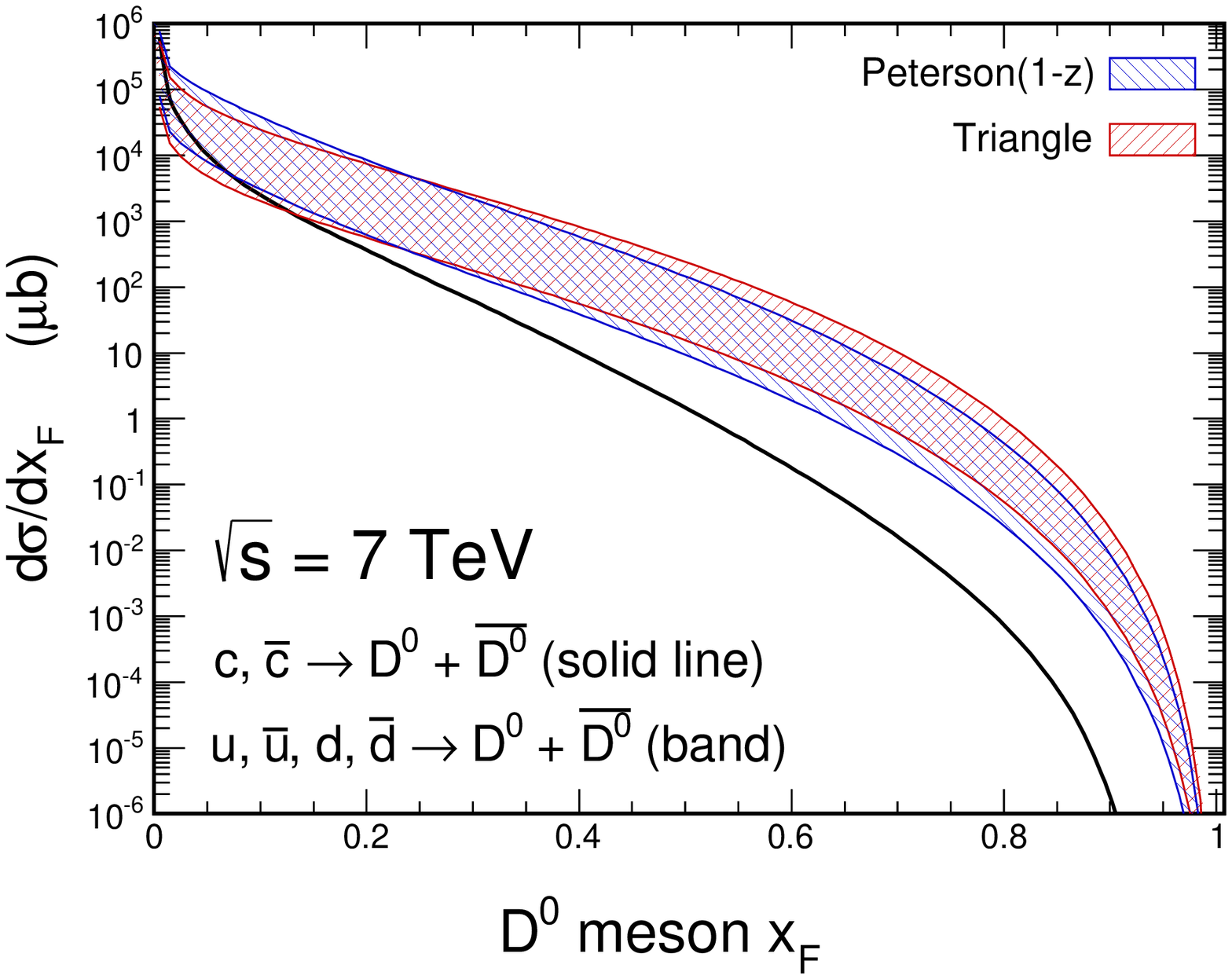}}
\end{minipage}
\begin{minipage}{0.47\textwidth}
  \centerline{\includegraphics[width=1.0\textwidth]{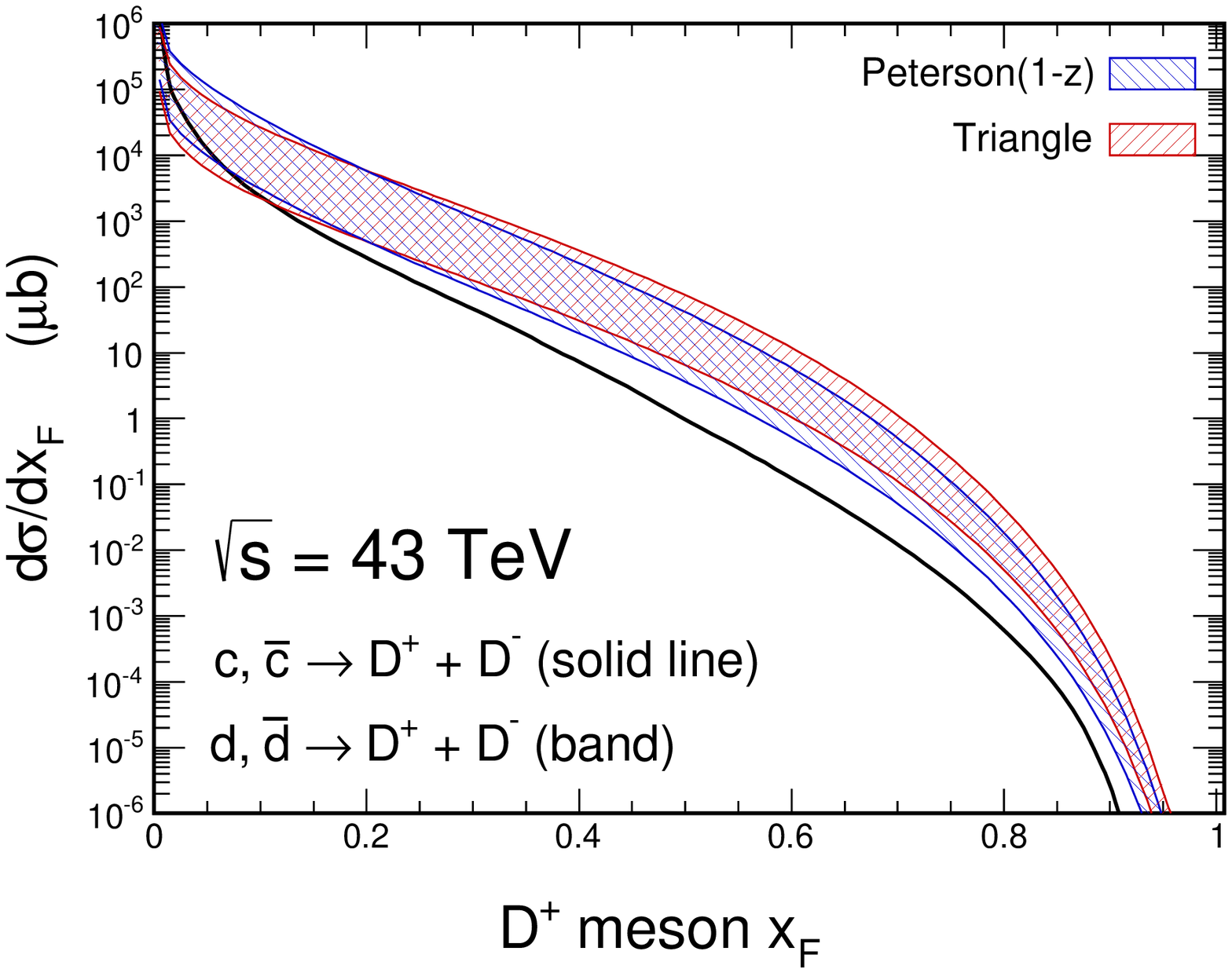}}
\end{minipage}
\hspace{0.5cm}
\begin{minipage}{0.47\textwidth}
  \centerline{\includegraphics[width=1.0\textwidth]{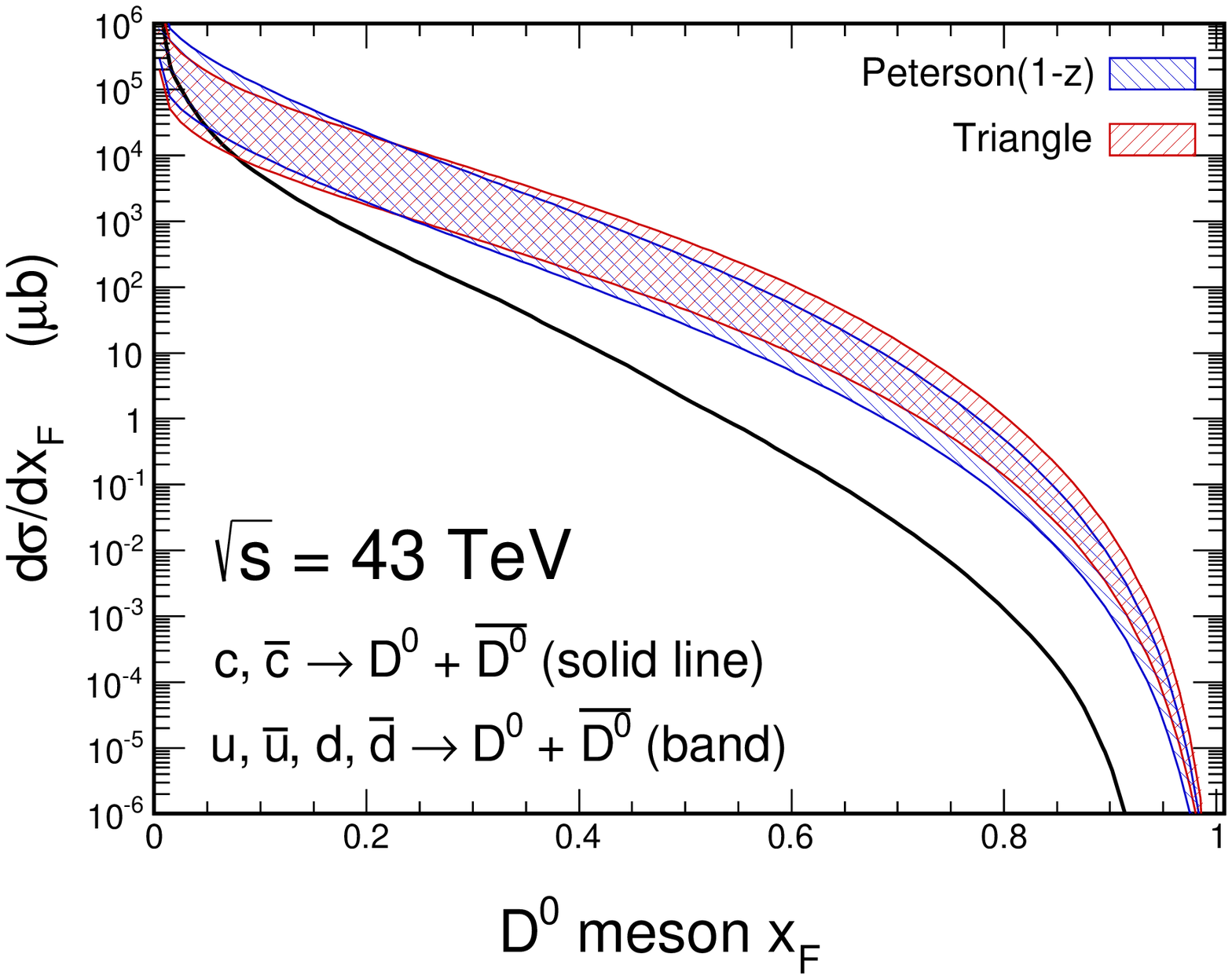}}
\end{minipage}
  \caption{
\small Distribution in $x_F$ for charged $D^+$+$D^-$ (left panel) 
and neutral $D^0$+${\bar D}^0$ (right panel) $D$ mesons from conventional 
(solid lines) and subleading (shaded bands) mechanisms.
The top panels are for $\sqrt{s}$ = 7 TeV and the bottom panels are for
$\sqrt{s}$ = 43 TeV.
}
\label{fig:dsig_dxf_D_mesons}
\end{figure}

In Fig.~\ref{fig:dsig_dxf_D_mesons} we compare 
the conventional contribution corresponding to $c \to D$ fragmentation and
the subleading one corresponding to $q \to D$ fragmentation,
for the sum of $D^{+}+D^{-}$ (left panels) and $D^{0}+\bar{D^{0}}$ 
(right panels) mesons.
While at small $x_F$ the conventional contribution dominates,
at large $x_F$ the situation is reversed.

\section{Conclusions}

We have discussed asymmetry in production
of $D^+$ and $D^-$ mesons in proton-proton collisions
as described in our recent original paper \cite{MS2018}, where
for a first time we tried to understand whether the asymmetry 
observed by the LHCb collaboration can be understood within parton 
fragmentation picture, including light quark and antiquark fragmentation
functions.

Very small unfavoured fragmentation functions are sufficient
to describe the LHCb data. The details depend however on functional form used.
The corresponding fragmentation probability for $q/{\bar q} \to D$ 
is of the order of a fraction of 1\%.
In \cite{MS2018} we showed predictions for similar asymmetry 
for neutral $D$ mesons.

We predicted large contribution of the light quark/antiquark
fragmentation to $D$ mesons at large 
$x_F$, which significantly exeeds the conventional $c/{\bar c} \to D$ 
contribution.

We calculated also the asymmetries for much lower energies
($\sqrt{s}$ = 20 -- 100 GeV), relevant for possible measurements
in a near future.
Much larger asymmetries were predicted, compared to those measured
by the LHCb collaboration. 
The asymmetries are associated with an increased production of charm in
the $q/{\bar q}$ initiated hadronization.
We quantified this effect by discussing corresponding asymmetries
and rapidity distributions.
The corresponding measurements at fixed target LHCb, RHIC,  
and at SPS (NA61-SHINE) \cite{NA61}
would allow to pin down the discussed here mechanisms.
Especially the SPS experiment could/should observe an enhanced
production of $D$ mesons. 

Systematic studies of $D/\bar D$ asymmetries at low energies may be
paradoxically important to understand the high-energy
prompt component of the atmospheric neutrino flux.
The predicted large contributions of $D$ mesons at large $x_F$ may have 
important consequences for prompt neutrino flux at large neutrino energies, 
relevant for the IceCube measurements.
We found that the contribution of the unfavoured fragmentation may be
more important than the conventional one for large
neutrino/antineutrino energies $E_{\nu} >$ 10$^{5}$ GeV.

{\bf Acknowledgments}

This study was partially
supported by the Polish National Science Center grant
DEC-2014/15/B/ST2/02528 and by the Center for Innovation and
Transfer of Natural Sciences and Engineering Knowledge in Rzesz{\'o}w.



\begin{thebibliography}{100}

\bibitem{IceCube_flux}
M.G. Aartsen et al.(IceCube collaboration) Astrophys. J.
{\bf 833} (2016) 3.

\bibitem{GMPS2017}
  V.~P.~Goncalves, R.~Maciu{\l}a, R.~Pasechnik and A.~Szczurek,
  Phys.\ Rev.\ D {\bf 96}, no. 9, 094026 (2017)
  [arXiv:1708.03775 [hep-ph]].

\bibitem{LHCb_asymmetry} 
  R.~Aaij {\it et al.} [LHCb Collaboration],
  Phys.\ Lett.\ B {\bf 718}, 902 (2013).

\bibitem{MS2018}
R. Maciu{\l}a and A. Szczurek,
arXiv:1711.08616, in print in Phys. Rev. {\bf D}.

\bibitem{SPS}
  T.~Anticic {\it et al.} [NA49 Collaboration],
  Eur.\ Phys.\ J.\ C {\bf 68}, 1 (2010).



\bibitem{LHCb_fixedtarget}
The LHCb collaboration, LHCb-CONF-2017-001.

\bibitem{FONLL}
  M.~Cacciari, M.~Greco and P.~Nason,
  J. High Energy Phys. {\bf 05} (1998) 007;\\
  M.~Cacciari, S.~Frixione and P.~Nason,
  J. High Energy Phys. {\bf 03} (2001) 006.

\bibitem{AguilarBenitez:1988sb} 
  M.~Aguilar-Benitez {\it et al.} [LEBC-EHS Collaboration],
  Z.\ Phys.\ C {\bf 40}, (1988) 321.

\bibitem{Ammar:1988ta} 
  R.~Ammar {\it et al.},
  Phys.\ Rev.\ Lett.\  {\bf 61}, 2185 (1988).

\bibitem{NA61}
Y. Ali, P. Staszel, A. Marcinek, J. Brzychczyk and R. P{\l}aneta,
Acta Phys. Polon. {\bf B41} (2013) 2019.

\end{thebibliography}
\end{document}